\documentclass[sigconf]{acmart}
\usepackage{booktabs}
\pdfoutput=1
\renewcommand\footnotetextcopyrightpermission[1]{} % removes footnote with conference information in first column
\begin{document}

\title{WorkerRep: Immutable Reputation System For Crowdsourcing Platform Based on Blockchain}

\author{Gurpriya Kaur Bhatia}

\affiliation{%
  \institution{Indraprastha Institute of Information Technology, Delhi	}
  }
\email{gurpriya16021@iiitd.ac.in}
% \orcid{1234-5678-9012}
%
\author{Shubham Gupta}
% \authornotemark[1]
\affiliation{%
  \institution{Indraprastha Institute of Information Technology, Delhi	}
  }
\email{shubhamg@iiitd.ac.in}

\author{Alpana Dubey}
\affiliation{%
  \institution{Accenture Labs, Bangalore}
}
\email{alpana.a.dubey@accenture.com}

\author{Ponnurangam Kumaraguru}
\affiliation{%
  \institution{Indraprastha Institute of Information Technology, Delhi	}
  }
 \email{pk@iiitd.ac.in}

\begin{abstract}
 Crowdsourcing is a process wherein an individual
or an organization utilizes the talent pool present over the Internet
to accomplish their task. The existing crowdsourcing platforms
and their reputation computation are centralized and hence prone to various attacks or malicious manipulation of the data by the central entity. A few distributed crowdsourcing platforms have been proposed but they lack a robust reputation mechanism. So we propose a decentralized
crowdsourcing platform having an immutable reputation mechanism
to tackle these problems. It is built on top of Ethereum network and does not require the user to trust a third party for a non-malicious experience. It also
utilizes IOTA`s consensus mechanism which reduces the cost for
task evaluation significantly.
\end{abstract}


\maketitle

\section{Introduction}
Jeff Howe first coined the term crowdsourcing, describing
it as "the act of a company or institution taking a function
once performed by employees and outsourcing it to an
undefined network of people in the form of an open call"~\cite{CSDef}. \par
\begin{figure}
    \includegraphics[width=\linewidth]{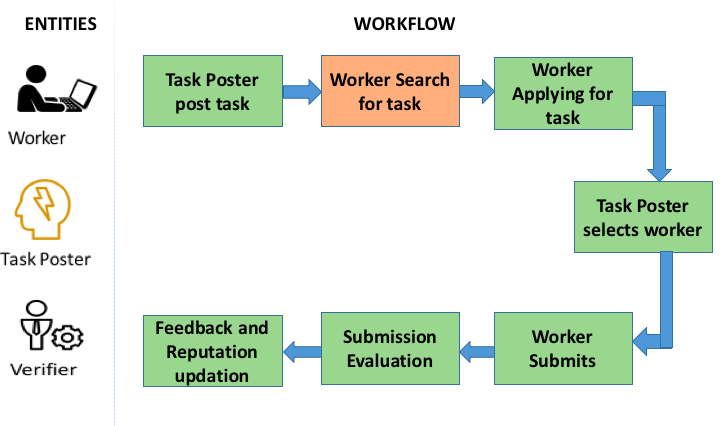}
    \caption{Entities and Various steps of Crowdsourcing platform}
    \centering
    \label{fig:steps}
\end{figure}

Several crowdsourcing platforms have emerged in the past
decade such as Amazon Mechanical Turk~\cite{mturk}, Upwork~\cite{upwork}, Topcoder \cite{topcoder}. These
platforms offer numerous advantages such as reduced cost,
better quality, and lower task completion time. Because of these advantages,
tasks ranging from as simple as data annotation to as complex
as software development are being crowdsourced. In general,
there are two types of crowdsourcing platforms: (1) hiring
based, in which the workers apply for a task. After the
application process is over the task poster chooses workers
from the set of applicants, to work on the task. The number
of workers depends upon the requirement of the task poster.
Only the selected workers are eligible for the reward, upon
successful completion. (2) Competition based platform, where there is no worker selection initially. 
Any number of workers can work on a task. But the reward
is awarded to the best one which is generally decided by the
task poster. In this paper, we focus on hiring based crowdsourcing platform which comprises of three entities,
namely the task poster who needs to get a task completed,
the worker who takes up the task to complete it and the
verifier checks for the accuracy, quality, etc. for the task completed by the worker. The entities involved along with the basic workflow followed by the platform is shown in Figure \ref{fig:steps}. \par
Majority of the
existing crowdsourcing platform such as Upwork, Topcode,
AMT, etc. follow a centralized management approach; that
is, they have a central authority through which all the
activities are managed. Here, central authority refers to the
people or the organization managing the platform. The
central authority charges a fee as a part of getting a task done
over the platform. Also, there is an assumption that
the central authority is trustworthy, which always needs to be correct\cite{hoffman2009survey}. There can be attacks on central authority either by outsiders or insiders who can maliciously manipulate data\cite{hoffman2009survey}. Moreover, there might be privacy concerns to the
user upon how his/her information will be used. In some
instances, user certification authorities have been shown to turned unreliable~\cite{zeilinger2018digital}.  Some crowdsourcing platforms require
the task poster to deposit the reward amount before beginning with the tasks. This gives central authority an undue
opportunity of deciding how it can use the money. In the worst
case, they may take away the money without users consent~\cite{soska2015measuring} or may not reward
the worker suitably on successful completion of the task.\par 
Even if we assume that central authority is trusted, there is
a need to establish trust among the task poster and the worker
for the success of the platform ~\cite{dwarakanath2016trustworthiness}. One of the important
aspect of trust comes from the reputation of the individual users of
the platform. The majority of existing studies have approached this from the perspective
of a centralized crowdsourcing platform ~\cite{klinger2011enabling,whiting2016crowd}. On such platforms, integrity and authenticity
of the reviews received and reputation scores computed
cannot be guaranteed as all the control resides with the central
entity and it is easy for them to manipulate it.

In this paper, we propose a blockchain based design philosophy for a crowdsourcing platform. Blockchain based design
principles model any system, that involves any form of
transactions among entities, as a decentralized transaction
management system where transactions are hashed cryptographically,
validated based on consensus, stored
by multiple entities subscribed to the system and are immutable. Bitcoin ~\cite{nakamoto2008bitcoin} is one of
the very famous application of blockchain. We show the
feasibility of the proposed system with Ethereum ~\cite{buterin2014next} based
implementation of the platform. The key contribution of our work is: (1) a decentralized crowdsourcing platform, (2) a
publically observable, robust and immutable reputation mechanism.\par

The rest of the paper is organized as follows. Section \ref{sec:relatedWork} briefs about the related work and section \ref{sec:model} describes the layout of our design. Section  \ref{sec:Process} explains how each step of the crowdsourcing process is carried out and \ref{sec:CompRep} explains how reputation on the platform is updated. The experiments performed and their result is mentioned in section \ref{sec:ex} and the analysis on how the platform overcomes various attacks is stated in section  \ref{sec:analysis}. Section \ref{sec:conFu} talks about the conclusion and future work.

\section{Related work} \label{sec:relatedWork}
With the rise in conventional workforce moving towards gig-economy \cite{2020Ref} there has been widespread conceptualization of it in different areas ranging from social science and humanities~\cite{ghezzi2018crowdsourcing} to software engineering~\cite{mao2017survey}. 

Various challenges of centralized crowdsourcing platform such as trust management~\cite{ye2015crowd,wu2015endorsement,whiting2016crowd,gaikwad2016boomerang,jagabathula2014reputation,dwarakanath2016trustworthiness,xie2015incentive}, incentive mechanism~\cite{xie2015incentive,zhang2015truthful,gao2015cost,kamar2012incentives}, quality control~\cite{allahbakhsh2013quality,zaidan2011crowdsourcing,oleson2011programmatic,baba2013statistical}, privacy and security~\cite{yang2015security,teo2018privacy,toch2014crowdsourcing} have been taken up in literature. 

On the other hand attacks and their corresponding defense on reputation systems have also been studied in the literature. Interconnecting reputation system and social networks ~\cite{golbeck2004accuracy} or clustering similar users ~\cite{tavakolifard2012taxonomy} has been suggested as a remedy to overcome the problem of cold start. Unfair rating attack can be avoided by using statistical measures to exclude such ratings ~\cite{dellarocas2000immunizing} or by weighting the rating by rater's reputation ~\cite{cornelli2002choosing} or comparing it with rating provided by some trusted agent apriori ~\cite{josang2009challenges}. 

Work on distributed reputation system by~\cite{anceaume2013privacy} wherein they have presented a privacy-preserving reputation mechanism by using a pseudonym for users rather than their actual identity and requires a certification authority for authentication. It also keeps discarding old reviews and provides no mechanism to prevent or handle an unfair rating. Zacharia in ~\cite{zacharia2000collaborative} proposes to use the most recent rating received by the user from the rater to overcome the collusion attack, but this might not be reflective of its past behavior and it would be easy to manipulate the rating.Authors in ~\cite{carboni2015feedback,schaub2016trustless,dennis2015rep}  propose reputation systems that leverages blockchain.
Carboni in ~\cite{carboni2015feedback} associates fees with every feedback the seller wants. This can reduce the review that is given for transactions that have not been performed but does not eliminate it. And might be burdensome for the seller. Authors in ~\cite{schaub2016trustless} have focused on computing the reputation of the seller on an e-commerce platform using reviews (rating as well as textual) that he/she received. Each review is bound to a transaction and reviewer anonymity can only be guaranteed if several transactions are to be reviewed which involves the seller in a given time. Ways of overcoming biased reviews have not been proposed. Dennis and Owen in their paper~\cite{dennis2015rep} use IP addresses to bind the identity of the user. Spoofing IP addresses is fairly easy now. To avoid collusion attack they are taking an average of the scores received by the worker. Since averaging is sensitive towards the outlier, it might reduce the effect of collusion but does not eliminate it.  \par
 Paper by Li, et al \cite{li2017crowdbc} is the most related to our work. They have proposed a crowdsourcing platform built on Ethereum and task assignment is done on a first come first serve basis which might result in poor quality and they have not touched much upon evaluation of the submission. They have proposed a reputation management scheme but have not looked upon various attacks that can occur. The other work that is similar to ours is of ~\cite{lu2018zebralancer}. They have proposed a crowdsourcing platform leveraging blockchain and that largely focuses on preserving privacy and anonymity of their users. However, they have not presented a reputation mechanism for their system and require users to have a new account for every transaction on the system. Both the above implementations require the central authority for authenticating user identity and issuing pseudonyms to the users.  

We propose a crowdsourcing platform that is decentralized and does not rely on third party for its core functionalities and it is built on top of Ethereum. It also has an added advantage of reduced cost as the task poster is not required to pay to any central agent to get his/her task done. We also try to prevent some of the various attacks on the reputation system, mentioned above, by building a stronger evaluation methodology.

\section{Our Model} \label{sec:model}
WorkerRep is modeled as a hiring based crowdsourcing
platform. There are mainly two entities on our platform (1) the task poster and (2) the worker. Unlike most of the crowdsourcing
platforms, where a task is evaluated by task poster, in
WorkerRep tasks are evaluated by peer workers present on the
platform. Based on their skills and reputation on the platform,
they are pseudo-randomly selected for task evaluation. To
evaluate is to judge if the submission by their co-worker satisfies the task requirements or not.

\subsection{Terminology} \label{term}

In this section, we describe various attacks and other terminologies
used in the paper.\par
\par \textit{Reputation} is the measure of how well the worker has
performed in the tasks that were previously assigned to him and how
well he evaluates the work done by others.
\textit{Initialization and Cold Start Problem} is faced by the new users on the platform to raise their reputation score initially. 
\par \textit{Sybil attack} is an attack where a malicious worker tries to
create multiple identities over the platform to gain influence
on the platform. Generally, It is done to carry out some of the
below-mentioned attacks.
\begin{itemize}
    \item \textit{Re-entry attack} carried out by creating a new identity
on the platform, leaving an identity with a bad reputation.
Generally, reputation lower than what is for a new worker
    \item \textit{Collusion attack} is when a group of workers tries to collude together to improve their own reputation or decrease the
the reputation of others.
    \item \textit{Ballot Stuffing} is when the worker tries to increase its
own reputation.
\end{itemize}

\par \textit{Unfair rating} attack is when the rater is biased towards the worker and does not give a truthful opinion about him. If it is biased in
a negative sense, that is rater tries to decrease the reputation
of the worker it is known as bad-mouthing.
\par \textit{Reciprocity} is when the worker reciprocates negatively for a
negative review that he receives.
\par \textit{Whitewashing} attack happens when either the worker knows
how to manipulate the reputation system or by re-entering the system.

\subsection{Layout}
\begin{figure} 
\includegraphics[scale=.35]{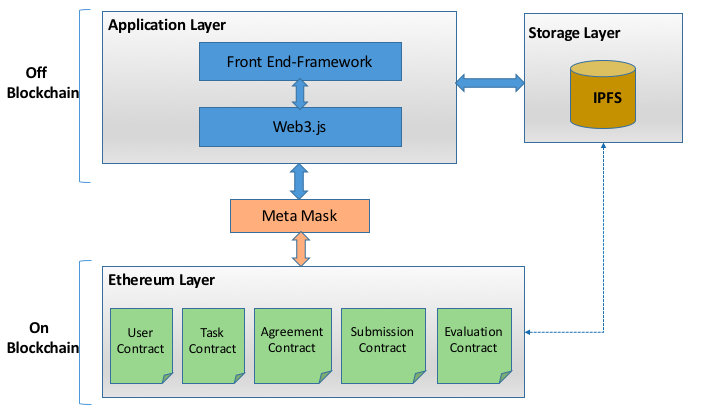}
\caption{Proposed crowdsourcing platform architecture}
\centering
\label{fig:arch}
\end{figure}

Since our proposed system is built on top of Ethereum
network, a smart contract is the most fundamental building
block for it. Our system has five kinds of smart contracts: 
(1) \textit{UserContract}: contains functionalities to create new users on the platform.  
(2) \textit{TaskContract}: it offers functionalities to create a new task as well as view the existing tasks on the platform. 
(3) \textit{AgreementContract}: creates an agreement between the task poster and the worker corresponding to a given task. 
(4) \textit{SubmissionContract}: the worker invokes the functionalities of this contract when he/she has finished the task he/she was assigned and is ready to submit. The basic functionality of this contract is to accept submission from the worker and the assigned evaluators for that submission. 
(5) \textit{EvaluationContract}: provides various functionalities to the evaluators of the tasks. It also computes and updates the reputation of workers based on the evaluation score received from the evaluators. The architecture of the system is shown in Figure \ref{fig:arch}.

To perform any action on the platform the user has to call the function of the contract corresponding to that action. If the function call results in a change of state of Ethereum Blockchain then the function call is treated as a transaction from the
one calling that function. Transactions on Ethereum network are cryptographically
signed using an asymmetric encryption algorithm to prevents
non-repudiation of the origin of the transaction as well as to
maintain the integrity of the data. Miners on the other hand, can decrypt the transaction
using the initiator's public key. After decrypting it they verify
the validity of the transaction and if found valid, these
transactions are appended to the public ledger. These transactions cost Ethers to the user and interactions with the Ethereum network are carried out by using MetaMask~\cite{metaRef}. MetaMask is an application that acts as a bridge between the browser and Ethereum. Using MetaMask saves users from installing the Full Ethereum node on their local system as it uses the full nodes from Infura~\cite{infuraRef}.  
In Figure \ref{fig:steps}, function calls corresponding to steps
colored in green are treated as transactions and carried out on Ethereum
blockchain and the ones in orange that are task search and apply for a task do not involve appending transaction to the Ethereum blockchain.

\section{Process} \label{sec:Process}
Subsections below define how each of the steps mentioned in Figure \ref{fig:steps} is performed on our platform. 

\subsection{User Registration}
Users register on the platform as a worker or task poster and specify his/her public key and the IPFS\cite{benet2014ipfs} hash of their information. InterPlanetary File System(IPFS) is a peer to peer storage systems that uses the hash of the data as it`s address. Ethereum wallet address is used to register on the platform.
A user can have multiple addresses and its corresponding public-private key pairs from a
single Ethereum wallet account. Since each user is linked
to a particular address and public-private key pair, the users created
from different key pairs will be different. So to avoid a Sybil
attack, we ask the user to pay some amount of Ethers while
registering. Authors\cite{douceur2002sybil} have shown that if there is a cost
associated with generating identity, the Sybil attack is greatly
reduced. Another way to avoid a Sybil attack is to associate an identity with the user, but it requires a third
party to validate the identity. We use the former approach to handling the Sybil attack due to the following reasons. Firstly, including a third party contradicts our architecture of a decentralized system.
Secondly, asking for a fee initially would keep spammers
away from our system, and only those who are interested will
be joining the platform. The Ethers taken as a fee from the user are stored in their account on the platform and can be returned when the user leaves the platform. The amount returned to the user depends on the reputation the user has at that point in time.

\subsection{Post Task}
To post a task, the task poster specifies the reward for the task, title, skills required and the IPFS hash of the task metadata and sends the transaction. Upon the transaction being mined, the task gets appended to the public ledger as a transaction
carried out by the task poster and to the list of available tasks in the \textit{TaskContract}. This step is incorporated within the Ethereum network so that further transactions happening for
this task such as assigning the task to the worker, submission
by the worker, etc., can be linked to it. It will help ease the verification process carried out by other task posters before assigning a worker to their task. 

\subsection{Task Search}
Once a task is posted, it is added to the list of available tasks for workers to perform. All the available tasks and their relevant details are public and available for view. Workers can apply filters on the collection of available tasks to aid them in choosing tasks. 

\subsection{Task Registration}
In this step, a worker can choose to apply to an available task posted on the platform. By applying, a worker can express his interest in that task. Details of workers who have applied for the tasks are visible to task posters and these can help the task poster in choosing worker(s) for their task. 

\subsection{Worker Selection and Task Assignment}
Selecting the most suitable worker from the applicants is a time-consuming process. The task poster can initially filter out workers based on their reputation score. From those selected after
filtering based on reputation, the task poster can go through
their profiles and evaluate them. Once the task poster has selected worker/s for his/her
task, he/she creates an agreement using the \textit{AgreementContract} specifying the
worker information within it. The agreement here is a binding structure between worker and task poster for a given task that can enforce the rules of the system associated with the task such as acceptance, submission, reward, and the updates to reputation.
 While creating the agreement, the task
poster is required to send the reward amount for that task,
to this agreement. This amount can only be withdrawn if the worker has not accepted the task and the task poster wants to cancel the task or can be awarded to the worker upon completion of the task. Since there is no central authority involved the money is kept safe in the contract and will be awarded to either of the two entities involved in the agreement based on the above-mentioned condition. Letting task posters send in money while creating assignments will help in avoiding task posters to
assign numerous tasks at a time, hence spamming the network
and also avoid task poster denying to pay to the worker upon
satisfactory task completion. After the agreement is successfully created, the selected worker is notified for
him/her to accept it and start working on it. These transactions
are recorded in the Ethereum.

\subsection{Task Acceptance}
Once the task poster assigns the worker to the task by creating an agreement, the worker is notified of the same. To accept the task, the worker is required to
send in some fixed amount of Ethers to that agreement.
The Ethers that are deposited by the worker will be refunded on successful completion of the task.
Upon receiving the Ethers from the worker the contract code will check
whether the address of the sender of those Ethers is
the same as worker address specified by the task poster while creating that
agreement. If it is the same, the task poster will be notified of the
acceptance and the worker can start working on the task else
the gas corresponding to that transaction will be consumed. The task can be accepted by a
specific date mentioned by the task poster, post which the
agreement will stand canceled automatically and all the Ethers
residing in that agreement will be sent to the creator of that agreement, which is to the task poster. Post this date the worker will not be able to accept the agreement. This is done so that the task poster does not keep waiting for an infinite time for the worker to accept the agreement and the task poster can receive the Ethers back and use them to assign the task to another worker. 
Just so the task poster does not
explicitly kill the agreement in the middle, we keep a check
if the worker has accepted the agreement. If he/she has
paid, then the contract cannot be killed by the task poster,
and it will only get terminated once the due date is reached.
Collecting Ethers from the worker when he is accepting the task works as an
assurance to the task poster that the worker would not leave
the task halfway. Since if he does not complete the task, the
Ethers submitted by the worker will be transferred to the task poster after the due date as compensation. And if he does the submission, the fees returned will depend on the completeness of the submission.

\subsection{Submission}
After the worker has completed the task, he can submit the
solution/submission by invoking the functions from the \textit{SubmissionContract}.
Initially, the worker hashes the submission using the Keccak-256 hashing scheme and sends it to the agreement before the due date. Hashing performed by the worker is not part of the Ethereum network. The worker performs if off blockchain, explicitly. Since any action that is performed on Ethereum is publically visible and we do not want to make the submission public. This hash received by the \textit{AgreementContract} serves two purposes (1) it helps the platform know that the worker has completed his/her work and is ready to get evaluated. (2) It helps prevent the worker from changing his submission upon knowing who his/her evaluator will be. So, upon receiving the hash, evaluators are pseudo-randomly chosen as discussed in the section below and their public keys are returned to the worker. Evaluators are not notified to the worker initially as it may introduce bias in his work if someone he knows becomes the evaluator of his task. \par
For each evaluator, the worker then encrypts the submission using the public key of that evaluator which is then encrypted by his/her own private key and is then stored on IPFS. The submission is not stored on blockchain due to (1) cost
involved to store large data on the blockchain network is high.
(2) Privacy concern, the task poster might be reluctant
to share the submission with the public. The hashes received by storing the encrypted submission for each individual evaluator are combined and then returned to the \textit{AgreementContract} by the worker. The first level of encryption provides confidentiality as nobody other than the evaluator having the corresponding private key can access the submission. The second level of encryption provides authentication of the worker. The case where evaluators explicitly share the submission with someone else can not be handled by our system. Encryption or decryption is not secure on blockchain since the keys become publically visible as in the case of hashing the submission as discussed above so, it is performed off blockchain at the user side. Storing the encrypted hash once for each worker consumes a lot of space. Instead, we can store the submission once over IPFS and perform the two-level encryption as described above to the hash received from the IPFS rather than performing it on the worker's submission itself. This would greatly reduce the space required on the IPFS as the number of evaluators increase but such a scheme can result in privacy and security issues as anybody with the hash can access the submission.

\subsection{Evaluation}
Evaluation on the existing crowdsourcing platforms is generally performed by the task poster itself. We propose a different model in which workers on the platform evaluate the submission of their peer workers and we also assume that the number of workers actively participating on the platform is sufficiently large. A worker evaluates a task when (1)
the worker wants to get his submission evaluated and (2)
the worker is willing to evaluate the
submission given he/she has a sufficiently high reputation. Our design requires the worker to evaluate \textit{y} other submissions before his submission gets evaluated. Here 
\textit{y} can depend on various factors such as the difficulty of the
task the worker has successfully submitted, the difficulty of the
tasks the worker is being assigned to evaluate, number of tasks that are yet to be evaluated, number of workers who are ready to evaluate and reputation
of the worker. One advantage of using this kind of evaluation model is that, as the number of nodes increases in
the network, the efficiency of submission evaluation increases. Such an evaluation model has been borrowed from IOTA's consensus model \cite{iota}. In the first case, various incentives for the worker to perform evaluation are that he does not have to pay fees to get his submission evaluated, he might learn new things that could be helpful for him in the future and there is an increase in reputation upon correct evaluation of peer's submission. In the second case, the incentive to evaluate others' submission is, to get their reputation increased and to learn. An increased reputation score in the latter case can easily be abused by a worker with a lower reputation as they would find this as a swift option to boost up their
reputation. To prevent this, a worker who has no submission that is pending for evaluation and has a low reputation score is not allowed to evaluate other's submissions. Only workers whose reputation is above a threshold are allowed to evaluate tasks if they have no pending submission to be evaluated. Further, this threshold can be altered dynamically to adjust to the current requirement of the number of evaluators. For our implementation, we have considered the threshold to be the average reputation of the workers on the platform.\par

There are two parts of review given by each evaluator
(1) The score, it comprises of two metrics, completeness and quality. Evaluator rates the submission, a value between [1, 100] on both the metrics. One means poor
performance and 100 means excellent performance. A range from 1 to 100 has been chosen because the floating-point division is not yet possible in solidity and such a range provides a balance between precision and memory requirements. (2) Textual review, a
brief explanation of the rating given by him/her.

Since peer workers are evaluating the task, there are chances that the evaluation may not be a true reflection of the work done by the worker due to unfair rating or collusion between the worker and the evaluators. To avoid it, we assign a total of \textit{x} evaluators to each task and take a consensus about their evaluation score to provide the worker with a fair evaluation. The incentive of increased reputation score as mentioned above might prompt the evaluator to randomly assign evaluation score to a submission. To decrease the effect of such random evaluation and unfair rating, the consensus among the evaluators is taken and the incentive of increased reputation is received by only those evaluators that are part of the consensus. Consensus mechanism is explained in section \ref{sec:CompRep}. 

The number of evaluators for a given submission can depend on various factors such as the complexity of the task which can be measured in terms of the duration required to complete the task, the number of workers available to evaluate the task which further depends on the number of people with the required skill and reputation, the number of workers who have to get their submission evaluated and the number of workers having the required reputation and willing to evaluate the submission. Once the number of evaluators is known, we next consider how they are chosen. For a given task, workers having the skill to evaluate that task are divided into slots based on their reputation score. The number of slots depends on the number of evaluators required for that task. Evaluators are then randomly chosen, one from each slot, assuming there always exists at least one worker in all the slots.  
The advantages of using this method are (1) allowing an evenly distributed pool of evaluators from different reputation scores to get better evaluations. (2) prevents workers who have made a submission and have a low reputation score from starving due to them not being allocated a task for evaluation, resulting in their task not being evaluated. (3) reduced probability of colluding workers being chosen together to evaluate the same task as successful collusion will require the colluding workers to firstly be in the list of eligible evaluators. Secondly, they would be required to be distributed in the reputation slots in such a way that would increase the chances of one of them getting selected from at least half of the slots. The collective effort and finances required to arrive at such a state far exceed the benefits that could be derived from it.

\section{Computing Reputation and Assigning Reward} \label{sec:CompRep}
There are two ways through which the reputation of the
worker gets updated (1) by completing a task and (2) by
evaluating other's submission. Initially, the reputation of the
worker is set to 1. Performance of the worker in every task that he/she 
has completed or evaluated, adds to their reputation on the
platform. The reputation of the worker based on his/her submission is dependent on the credibility of the evaluators and the score he/she receives for his/her submission. Likewise, when he/she evaluates a submission, then reputation depends on how close his/her score is to the score assigned to the worker.   

\subsection{Completing task}
As mentioned in section \ref{sec:Process} there are two criteria based on which the evaluators score the worker, which is completeness and quality.
Let \textit{$e_1$}, \textit{$e_2$}, ... , \textit{$e_n$} be the set of evaluators chosen pseudo-randomly and $c_{ik}$ and $q_{ik}$ be the \textit{completenessScore} and \textit{qualityScore} respectively given by the $i^{th}$ evaluator for $k^{th}$ submission. The mean of the \textit{completenessScore} and \textit{qualityScore}
suggested by the evaluators is denoted by $c_m$ and $q_m$ . Considering
$c_m$ and $q_m$ as a measure of \textit{completenessScore} and \textit{qualityScore} won’t be the correct as there might be malicious evaluators trying to collude or provide other workers with an unjust score. So to find and remove such malicious workers that are outliers among the evaluators, we compute the standard deviation
\textit{$c_s$} of the \textit{completenessScore} and $q_s$ as the standard deviation of \textit{qualityScore} specified by the evaluators. Evaluators whose score is beyond a certain threshold away from the mean in any of the two metrics, we tag him/her as an outlier. Remaining evaluators are said to form a consensus. 
The evaluation of the submission of the evaluators who are found to be outliers is not considered. This helps to deter evaluators from maliciously increasing or decreasing other worker`s reputation. If consensus is not reached, evaluators are reassigned and the above-mentioned steps are repeated. There can be better outlier detection methodology but due to the limited computational power provided by Ethereum, we stick to this methodology. 

Once the consensus is reached, the scores given by the evaluators on two different criteria are combined. Let $completnessScore_{jk}$ and $qualityScore_{jk}$ represent the consensus of the evaluator on the completeness and quality,respectively, of the submission \textit{k} made by worker \textit{j}. $completeScore_{jk}$ and $qualityScore_{jk}$ can be computed using equation \ref{completness} and \ref{quality}, respectively. 

\begin{equation}  \label{completness}
    completeScore_{jk} = \dfrac{ \sum_{i=0}^{ec} c_{ik} * r_i}{\sum_{i=0}^{ec} r_i} 
\end{equation}

\begin{equation}  \label{quality}
    qualityScore_{jk} = \dfrac{ \sum_{i=0}^{ec} q_{ik} * r_i}{\sum_{i=0}^{ec} r_i} 
\end{equation}

Here \textit{ec} represents the number of evaluators in consensus and $c_{ik}$ and $q_{ik}$ are the scores for completeness and quality respectively given by the $i^{th}$ evaluator for $k^{th}$ submission. Each of their scores is supported by the credibility of that evaluator. Credibility is measured in terms of the reputation of the evaluator on the platform and is represented as $r_i$ for $i^{th}$ evaluator.

Hence the final evaluation score ${finalScore}_{jk}$ assigned to the $ j^{th} $worker for a $k^{th}$ submission is the weighted average of score received in the two criteria. The weight assigned to these criteria can be decided by the task poster and sums up to 1. Suppose $w_c$ is the weight assigned to the $completeScore_{jk}$ and $w_q$ is the weight assigned to the $qualityScore_{jk}$. Then ${finalScore}_{jk}$ is calculated as : 

\begin{equation}  \label{score}
{finalScore}_{jk}= \frac{w_q * qualityScore_{jk}+ w_c * completeScore_{jk}}{w_q + w_c} 
\end{equation}
 
Upon computing ${finalScore}_{jk}$, it is then sent to the worker as a transaction. By sending a transaction to the worker we are trying to maintain a chain of reviews the worker has
ever got on the platform. This makes it easier for anybody
to prove the existing reputation of the worker. In this case, no central party can maliciously alter the
reputation score, and no worker can fake its reputation on the
platform. The score received by the worker is then added to the reputation of the worker.

The reputation is updated only if the worker makes a submission and it is not altered in the case when the worker accepts the task and does no submission. 

\subsection{Evaluating submission}
The other way of increasing reputation is by evaluating the submission made by peer workers. The score received by the evaluator upon evaluating a task depends on how far is its evaluation for quality and completeness for a given submission from the $completnessScore_{jk}$ and $qualityScore_{jk}$ received by the worker. Let the score received by evaluator \textit{i} upon evaluating submission \textit{k} be represented by $eScore_{ik}$. It can be computed using equation \ref{evalScore}

\begin{equation}  \label{evalScore}
    eScore_{ik} = \dfrac{200-\left|qualityScore_{jk} - q_{ik}\right| - \left|completeScore_{jk} - c_{ik}\right|}{2} 
\end{equation}

\subsection{Updating Reputation}
For every submission the worker does, he/she is required to evaluate \textit{y} other tasks represented by v . Let $\alpha$ be the weight assigned to the score that the worker receives upon evaluating other submissions to get his/her $l^{th}$ submission evaluated. And let (1- $\alpha$) be the weight assigned to the score that the worker receives for his $l^{th}$ submission. The value of $\alpha$ is taken to be 0.25 as authors\cite{de2014crowdgrader} found that giving 25 \% weightage for evaluating performance of their peers provided them sufficient motivation. Then the total reputation increased of the worker by doing submission and its correspoding evaluation is given by equation \ref{updateRep}
\begin{equation}  \label{updateRep}
    repScore{jk} = (1-\alpha){finalScore}_{jk} + \alpha\dfrac{\sum_{b=0}^{\left|v\right|} eScore_{jb}}{y}
\end{equation}

On the other hand, if the worker \textit{j} is willing to evaluate even though he has no pending submission to be evaluated, then the reputation score increased for that particular submission \textit{k}`s  evaluation is given by equation \ref{onlyEval}

\begin{equation}  \label{onlyEval}
    repScore_{jk} = \alpha * eScore_{jk}
\end{equation}

Both the above cases consider only those workers that are not outliers. But when the worker is an outlier, in the latter case the workers are decentralized as their reputation is decreased. The decrease is proportional to their $eScore_{jk}$ and is given by equation \ref{onlyEvalOutlier}

\begin{equation}  \label{onlyEvalOutlier}
    repScore_{jk} = -\alpha * eScore_{jk}
\end{equation}

For the former case, the worker is required to correctly evaluate the tasks before his task can get evaluated. Since evaluators are randomly chosen, the probability of him/her getting the next task early in time is low, hence delaying his submission to be evaluated. 

\subsection{Reward}
The reward received by the worker wholly depends on the ${finalScore}_{jk}$ he receives for his submission. If task reward as posted by the task porter for task \textit{t} is $taskReward_t$ then the reward received by the worker \textit{j} for submission \textit{k} corresponding to task \textit{t} represented as $reward_{jk}$ and can be computed using equation \ref{reward}

\begin{equation} \label{reward}
reward_{jk} = \dfrac{{finalScore}_{jk} * taskReward_t}{100}  
\end{equation}

Apart from the reward, the worker also gets back the fee (\textit{acceptanceFees}) that he/she paid while accepting the task. The fees returned ($feeReturned_{jk}$) to the worker depends on the completeness of the submission and is computed using equation \ref{fees} 

\begin{equation} \label{fees}
feeReturned_{jk} = \dfrac{completeScore_{jk} * acceptanceFees}{100}  
\end{equation}

The reward and the acceptance fee that remains are sent to the task poster.

\section{Experiments and Results} \label{sec:ex}
To perform experiments and test the performance, validity and other parameters of our system, we have developed an implementation on Ethereum blockchain using solidity which is a programming language used to develop smart contracts. The resulting implementation was organized into 5 different contracts with each smart contract handling a specific group of functions related to our system, details for which have already been described in Section \ref{sec:Process}.
 
% While our system is designed to be able to process tasks and workers involving various skills, our test implementation, however, assumes a single skill to be the focus of our application.

We also developed a front end web interface that makes it easy for our users to interact with our platform and also can hide the complexity behind the function calls. We have used web3.js, which is a collection of Ethereum JavaScript API libraries that enable our front end to connect to and talk to our smart contract on the Ethereum blockchain. For creating the user interfaces and views we have used Reactjs, a popular JavaScript library for creating user interfaces and Metamask has been used for signing transactions. While our system is not tied to any particular storage system, we have used IPFS in our implementation due to its decentralized and distributed nature. Our frontend interacts with IPFS via ipfs-api which is its JavaScript API and using this API allows us to add and retrieve files from IPFS.

When a worker registers on our platform, he or she is expected to provide an IPFS link (hash) to his/her profile or portfolio which contains information that would be helpful to task posters while choosing workers for their task. Also while registering, a worker is expected to provide a fee as a deposit which we have chosen in our implementation to be ether equivalent of \$5  (0.0118 ether at the time of writing).

Once solidity code is compiled, all the functions are converted into low-level assembly opcodes suitable for the Ethereum Virtual Machine 
(EVM). The gas cost or cost of executing each of these low-level operations is fixed and defined in the Ethereum yellow paper. So the gas cost for any function is a measure of the cost of its processing requirements and for any function or any other particular flow of control, the amount of gas required for its successful execution is constant and can be determined deterministically. Wallets such as Metamask also provides an estimate of gas cost for the execution of function based on the same idea.  

To gain a better idea of the transaction costs involved in executing the major functionalities of our application, we first obtain gas requirements and then estimate their cost by taking a price of 1 Gwei (Giga-Wei) per gas. It may be noted that 1 billion Gwei is equal to 1 ether and the price of 1 ether at the time of this experiment (Dec 2019) is \$144.30.

\begin{table}
% \centering
\caption {Gas requirement and cost of executing various functions of our smart contract}
\label{tab:dataset}
\begin{tabular}{ccl}

% \begin{tabular}{@{}l|c|c|c|c@{}}

\toprule
   Function           & Gas  & Cost estimate (\$) \\ \midrule

Create worker     &229,786 &0.0333    \\
Create taskposter     &228,410 &0.0331\\
Post task with fees      &250,502 &0.0363 \\
Create agreement     &198,134 &0.0287 \\
Accept agreement     &49,729 &0.0072 \\
Submit hash      &114,068 &0.0165 \\
Assign evaluators  &328,702 &0.0477 \\     
First evaluation submit &133,073 &0.0193 \\     
Second evaluation submit     &105,620 &0.0153 \\
Third evaluation submit    &274,360 &0.0398 \\
Become evaluator     &47,878 &0.0069 \\ \bottomrule
\end{tabular}
\end{table}

Using the data from Table 1, we calculate the total cost of a task being posted to it being evaluated on our system. The total gas required for this is calculated to be 1,502,066 which translates to \$ 0.2178. The cost of registering as a worker or a task poster has not been included in this since those costs are incurred only once.

During the course of the development of our smart contracts, various options were explored for testing the smart contracts. One method is to use software such as ganache-cli that provide a local version of Ethereum blockchain for testing. Another popular option is to use test networks such as Rinkeby or Ropsten. These test networks work in a manner identical to the Ethereum main net and only differ in the fact that ether on these networks do not actually have any value and they are only used for testing. 
% since square root operation is not directly available we have used (Taylor series method + cite) to obtain square roots to calculate standard deviations from the calculated variance for the scores provided by the evaluators.

%The operation of assigning the evaluator is the most expensive since it requires traversing through the entire list of workers and filling the various reputation slots. For our work, we have taken the number of evaluators to be 3 and the number of tasks to be evaluated by the worker also to be 3 since it provides a balance between the time taken for a complete evaluation cycle and the   \textcolor{red}{something }.  

\section{Analysis of the Design} \label{sec:analysis}
In this section we explain how various objectives of our system are fulfilled. 
\subsection{Privacy and Anonymity}
Our system maintains the privacy of task submissions by using asymmetric-key encryption techniques. Apart from the evaluators and task poster, no other users will be able to view the submitted work. Rather than providing anonymity our platform does provide pseudo-anonymity to its user by not linking an account with its personally identifiable information. Pseudo anonymity is inherited from Blockchain. \cite{lu2018zebralancer} provides anonymity by not linking transactions from two different tasks. But this would fail to build trust upon each other on the platform. As the actions performed by a user in the past would not be known to the other users making it easier for him/her to mislead others on the platform.

\subsection{Decentralization}
There is no central authority in charge of running or maintain data on the platform. All functions carried out by the users on the platform are validated by the miners on the Ethereum network. There are incentives for the user to show good behavior and a disincentive for malicious activities.

\subsection{Robustness against various attacks}
This section talks about how our design helps prevent or sometimes avoid various attacks as mentioned in \ref{term}. 
\begin{itemize}
    \item \textit{Unfair rating attack/ Bad mouthing} 
        \begin{itemize}
        \item A consensus among various evaluators is being taken. The consensus mechanism involves removing outlier/s. The outlier here means the evaluator whose evaluation score is not in line with the majority of the evaluators. The final rating does not consider the score given by the outlier/s.
        \item The reputation of the evaluator is decreased or he/she is given another task for evaluation if he/she is found to be an outlier. This acts as a major deterrent for giving a rating that is not consistent with the work that is done by the peer worker. 

        \end{itemize}
    \item \textit{Reciprocity} : Firstly the chances of getting selected as an evaluator for that worker are less, given several evaluators and secondly outliers are being dealt with while computing the final score for the worker, it will be tough for any worker to reciprocate.  
    \item \textit{Ballot Stuffing and collusion Attack} :  these are prevented by providing a robust evaluation methodology as described in the evaluation in section \ref{sec:Process}.
    
%     Each of the reputation transaction received by the worker is from a workerTaskPoster smart contract. The value received in the reputation transaction is in turn calculated from the evaluation score transactions   received by workerTaskPoster contract from the evaluators. Since the evaluators are pseudo-randomly chosen it is hard for users to collude and for a worker to increase his own reputation. ( we will have to consider some assumption or some probability with which this attack is not possible) \textcolor{red}{redo this }

    \item \textit{White washing/ Re-entry / Sybil} : 
    This is prevented by adding an initial entry fee to the platform. The amount returned to the worker when he/she leaves the platform depends on their reputation at that time.
    
    \item \textit{Cold Start Problem} : 
    we have some low paying tasks on the platform that can be used as a starting point for new workers.

\end{itemize}

\section{Conclusion and Future work} \label{sec:conFu}
In this work, we propose a distributed and secure crowdsourcing platform with a robust reputation management scheme. The existing centralized schemes are susceptible to attacks on central servers or misuse by the central authority. Using the blockchain based technique we build a platform that is less expensive than the existing, does not rely on any third party and overcomes malicious manipulation of reputation and various other attacks on the reputation system. It also provides traceability, prevents unvalidated modification of data and gives a fair share of reward for the worker and compensation for the task poster. The reputation score provided by the system can be dependable due to immutability inherited by the system from the blockchain. And we also provide Proof of concept of our design by building the system on Ethereum test net.\par
However, further discussion is required on some of the points and we keep them as our future work. Firstly, a more effective technique of assigning submissions to a worker to evaluate can be developed that can relate to the previous tasks that the worker completes and can depend on how recently the worker has completed an evaluation. Secondly, there is a delay of 2.4 minutes on average\cite{ethNet} while performing experiments between the creation of a transaction and its confirmation by other nodes. This delay is expected to reduce with the upcoming transaction scalability updates in Ethereum\cite{scaleRef}. Thirdly, doing computation on Ethereum blockchain is very expensive and can be reduced by using computational oracles that perform the computations off-blockchain instead of doing it on blockchain but this is done at the cost of incorporating centralization. Lastly, our design was based on peers evaluating completed tasks and the task poster did not play any role in this step. Possible modifications can be explored that allow the task poster to have a reputation score and for him/her to play a  greater role in the evaluation step.

\bibliographystyle{ACM-Reference-Format}
\bibliography{mainFile}

\end{document}